\documentstyle[preprint,aps,prd]{revtex}

\begin{document}

\draft

\title{Classical behaviour after a phase transition}

\author{F. \ C.\ Lombardo$^{1}$ \thanks{f.lombardo@ic.ac.uk},
F. \ D.\ Mazzitelli$^{2}$\thanks{fmazzi@df.uba.ar}, and R.\ J.\
Rivers $^{1}$\thanks{r.rivers@ic.ac.uk}}

\address{{\it
$^1$ Theoretical Physics Group; Blackett Laboratory, Imperial
College,
London SW7 2BZ\\
$^2$ Departamento de F\'\i sica, Facultad de Ciencias Exactas y Naturales\\
Universidad de Buenos Aires - Ciudad Universitaria,
Pabell\' on I\\
1428 Buenos Aires, Argentina}}

\maketitle

\begin{abstract}
We analyze the onset of classical behaviour after a second-order
phase transition by considering a scalar field theory in which the
system-field interacts  with its environment, represented both by
further fields and by its own short-wavelength modes.  Within our
approximations we see that the long-wavelength modes have become
classical by the time that the transition has been first
implemented (the spinodal time).
\end{abstract}

\vskip2pc 
\narrowtext

\vskip2pc Cosmology and particle physics suggest strongly that
phase transitions have occurred in the early Universe, in
particular at the grand unified and electroweak scales\cite{Kolb}.

An analysis of phase transitions in quantum field theory that
takes the non-equilibrium nature of the dynamics into account from
first  principles is very difficult, and has only begun to be
addressed. In particular, the naive picture of a classical order
parameter (inflaton or Higgs) field $\phi$ rolling down an
adiabatic effective potential, that was once a mainstay of
cosmological field theory modelling, has been shown to be suspect
\cite{cormier}. Alternatively, the suggestion by
Kibble\cite{Kibble} that, while a non-adiabatic approach is
crucial, causality alone can set saturated classical bounds on
time and distance scales during a transition, has been shown to be
only partly true. The issue of how a quantum system
evolves into the classical theory has been addressed in
Refs.\cite{salman,mottola}. For some models, it has been shown
that classicality emerges as a consequence of profuse particle
creation, whereby a non-perturbatively large occupation number of
long-wavelength particles produces, on average, a diagonal density
matrix. This dephasing effect occurs at late times. Here we will
consider a model of an explicitly {\it open} system, in which
classicality is an early time event, induced by the environment.
The result is completely different from that of
Refs.\cite{salman,mottola}.

There are several time scales which are relevant for the
description of the onset of a transition. If the quench is fast,
the initial stages of a scalar transition can be described by a
free field theory with inverted potential, $({\rm mass})^2 <0$.
This description is valid until the field wave functional explores
the ground states of the potential at the spinodal time $t_{\rm
sp}$. Specifically, the field ordering after the transition is due
to the growth in amplitude of its unstable long-wavelength modes.
In consequence, the short-wavelength modes of the field, together
with all the other fields $\chi_{\rm a}$ with which the $\phi$
inescapably interacts, form an environment whose coarse-graining
makes the system-field classical\cite{lombmazz}. As a result,
there is an additional time scale associated with the environment,
the decoherence time $t_{\rm D}$ \cite{zurek84,habib}. Once
$t>t_{\rm D}$ the order parameter becomes a classical entity
\cite{diana}.

In this letter we consider a simple model of a scalar
order-parameter field $\phi$, whose ${\cal Z}_2$ symmetry is
broken by a double-well potential. Specifically, we take the
simplest classical action with scalar environmental fields
$\chi_{\rm a}$ ($\mu^2, m_{\rm a}^2 >0$)
\begin{equation}
S[\phi , \chi ] = S_{\rm syst}[\phi ] + S_{\rm env}[\chi ] +
S_{\rm int}[\phi ,\chi ], \label{S}
\end{equation}
\begin{eqnarray}
&&S_{\rm syst}[\phi ] = \int d^4x\left\{ {1\over{2}}\partial_{\mu}
\phi\partial^{\mu} \phi + {1\over{2}}\mu^2 \phi^2 -
{\lambda\over{4!}}\phi^4\right\}, \nonumber \\
&&S_{\rm env}[\chi_{\rm a} ] = \sum_{\rm a=1}^N\int d^4x\left\{
{1\over{2}}\partial_{\mu}\chi_{\rm a}
\partial^{\mu}
\chi_{\rm a} - {1\over{2}} m_{\rm a}^2 \chi^2_{\rm a}\right\},
\nonumber
\\
&&S_{\rm int}[\phi ,\chi ] = - \sum_{\rm a=1}^N\frac{g_{\rm a}
}{8} \int d^4x \phi^2 (x) \chi^2_{\rm a} (x). \nonumber
\end{eqnarray}
As we have already observed, the separation in (\ref{S}) is not
yet the separation between the ultimately classical 'system' and
its 'environment' since the short-wavelength modes of the
$\phi$-field never become classical, and must be treated as part
of the decohering environment. However, to demonstrate how an
environment renders the order-parameter field classical we
consider the simpler case in which the environment is taken to be
composed of the $N$ fields $\chi_{\rm a}$ alone. Since the effect
of environmental sources is additive, at our level of
approximation, the $\chi_{\rm a}$ fields alone give us an {\it
upper} bound on $t_D$.

For weak couplings $\lambda ,~ g_{\rm a}\ll 1$ and comparable
masses $m_{\rm a}\simeq \mu$ we shall find that $t_D$ is  shorter
than the spinodal time $t_{\rm sp}$, defined as the time for which
$\langle \phi^2\rangle_t \sim \eta^2= 6\mu^2/\lambda$, the ground
states of the system.  In consequence, by the time that the field
is ordered it can be taken to be classical.

 The model has a continuous transition at a
temperature $T_{\rm c}$. The environmental fields $\chi_{\rm a}$
reduce $T_{\rm c}$ and, in order that $T_{\rm
c}^2=\mu^2/12(\lambda + \sum g_{\rm a})\gg \mu^2$, we must take
$\lambda + \sum g_{\rm a} \ll 1$. For order of magnitude
estimations it is sufficient to take identical $g_{\rm a} =\bar
g/\sqrt{N}$. We will also assume $\bar g\simeq\lambda$. For
one-loop consistency  in our subsequent calculation we assume that
$N\gg 1$. The effect of many weakly coupled environmental fields
is that they act on the system field without it being able to act
back on them.  For example, the dominant hard loop contribution of
the $\phi$-field to the $\chi_{\rm a}$ thermal masses is $\delta
m^2_T = O (\bar{g} T^2_{\rm c}/\sqrt{ N}) = O(\mu^2/N)\ll \mu^2$.
Similarly, the two-loop (setting sun) diagram, which is the first
to contribute to the discontinuity of the $\chi$-field propagator,
is of magnitude $\bar{g}^2 T_{\rm c}^2/N  =
O(\bar{g}\mu^2/N^{3/2})\ll\delta m^2_T$, in turn. That is, the
effect of the thermal bath on the propagation of the environmental
$\chi$-fields is ignorable. We stress that this is not a Hartree
or large-N approximation of the type that, to date, has been the
major way to proceed\cite{mottola,boya} for a {\it closed} system.

We shall assume that the initial states of the system and
environment are both thermal, at a temperature $T_{0}>T_{\rm c}$.
We then imagine a change in the global environment (e.g. expansion
in the early universe) that can be characterised by a change in
temperature from $T_0$ to $T_{\rm f}<T_{\rm c}$. The relevant
object is the reduced density matrix $\rho_{{\rm
r}}[\phi^+,\phi^-,t]=\langle\phi^+\vert {\hat\rho}_{\rm r}
(t)\vert \phi^- \rangle$. It describes the evolution of the system
under the influence of the environment, defined by
\[
\rho_{{\rm r}}[\phi^+,\phi^-,t] = \int {\cal D}\chi_{\rm a} ~
\rho[\phi^+,\chi_{\rm a} ,\phi^-,\chi_{\rm a} ,t],
\]
where $\rho[\phi^+,\chi^+_{\rm a} ,\phi^-,\chi^-_{\rm a},t]=
\langle\phi^+ \chi^+_{\rm a}\vert {\hat\rho}(t) \vert \phi^-
\chi^-_{\rm a}\rangle$ is the full density matrix.  The
environment will have the effect of making the system effectively
classical once $\rho_{\rm r}(t)$ is, approximately, diagonal in
the field configuration basis. Quantum interference can then be
ignored and we obtain a classical probability distribution from
the diagonal part of $\rho_{\rm r}(t)$, or equivalently, by means
of the reduced Wigner functional, which is positive definite after
the decoherence time (see Ref.\cite{diana} for an explicit
demonstration of this for a toy model). This behaviour  is
essentially different from the dephasing effects \cite{mottola}.
Our onset of classical behaviour is an early-time event which,
beneficially, allows us to use perturbation theory.

 Assuming that
the initial full density matrix can be factorised, the temporal
evolution of the reduced one is given by
\[\rho_{\rm r}[\phi_{{\rm
f}}^+,\phi_{{\rm f}}^-,t]= \int d\phi_{{\rm i}}^+ \int d\phi_{{\rm
i}}^- J_{\rm r}[\phi_{{\rm f}}^+,\phi_{{\rm f}}^-,t\vert
\phi_{{\rm i}}^+,\phi_{{\rm i}}^-,t_0] ~~\rho_{\rm r}[\phi_{{\rm
i}}^+ \phi_{{\rm i}}^-,t_0],
\]
where $J_{\rm r}$ is the reduced evolution operator

\begin{equation}
J_{\rm r}[\phi_{{\rm f}}^+,\phi_{{\rm f}}^-,t\vert \phi_{{\rm
i}}^+,\phi_{{\rm i}}^-,t_0] =\int_{\phi_{{\rm i}}^+}^{\phi_{{\rm
f}}^+} {\cal D}\phi^+ \int_{\phi_{{\rm i}}^-}^{\phi_{{\rm f}}^-}
{\cal D}\phi^- ~e^{i\{S[\phi^+] - S[\phi^-]\}} F[\phi^+,\phi^-].
\label{evolred}
\end{equation}
The Feynman-Vernon \cite{feynver} influence functional $F[\phi^+,
\phi^-]$ is defined as

\begin{eqnarray}
F[\phi^+,\phi^-] &=& \int d\chi^+_{{\rm a i}} \int d\chi^-_{{\rm a
i}} ~ \rho_{\chi}[\chi_{{\rm a i}}^+,\chi_{{\rm a i}}^-,t_0] \int
d\chi_{{\rm a f}} \int_{\chi^+_{{\rm a i}}}^{\chi_{{\rm a
f}}}{\cal D}\chi^+_{\rm a} \int_{\chi^-_{{\rm  a i}}}^{\chi_{{\rm
a f}}}
{\cal D}\chi^-_{\rm a}\nonumber \\
&\times & \exp{\left(i \{S[\chi^+_{\rm a} ]+S_{{\rm int}}
[\phi^+,\chi^+_{\rm a} ]-i\{S[\chi^-_{\rm a}] + S_{{\rm
int}}[\phi^-,\chi^-_{\rm a}]\} \right)}. \nonumber
\end{eqnarray}

Given our thermal initial conditions it is not the case that the
full density matrix has $\phi$ and $\chi$ fields uncorrelated
initially, since it is the interactions between them that leads to
the restoration of symmetry at high temperatures. Rather, on
incorporating the hard thermal loop `tadpole' diagrams of the
$\chi$ (and $\phi$) fields in the $\phi$ mass term leads to the
effective action for $\phi$ quasiparticles,
\[
S^{\rm eff}_{\rm syst}[\phi ] = \int d^4x\left\{
{1\over{2}}\partial_{\mu} \phi\partial^{\mu} \phi - {1\over{2}}
m_{\phi}^2(T_0) \phi^2 - {\lambda\over{4!}}\phi^4\right\}
\]
where $m_{\phi}^2(T_0)=-\mu^2 (1-T_0^2/T_{\rm c}^2)>0$. As a
result, we can take an initial factorised density matrix of the
form ${\hat\rho}[T_0] = {\hat\rho}_{\phi}[T_0]
{\hat\rho}_{\chi}[T_0]$, where ${\hat\rho}_{\phi}[T_0]$ is
determined by the quadratic part of $S^{\rm eff}_{\rm syst}[\phi
]$ and ${\hat\rho}_{\chi}[T_0]$ by $S[\chi_{\rm a} ]$.

Provided the change in temperature is not too slow the exponential
instabilities of the $\phi$-field grow so fast that the field has
populated the degenerate vacua well before the temperature has
dropped to $T_{\rm f}$. As $T_{\rm c}$ has no particular
significance for the environment field, for these early times we
can keep the temperature of the environment fixed at $T_{\chi} =
T_0={\cal O}(T_{\rm c})$. Since it is the system-field $\phi$
field whose behaviour changes dramatically on taking $T_{\phi}$
through $T_{\rm c}$, we adopt an {\it instantaneous} quench for
$T_{\phi}$ from $T_0$ to $T_{\rm f}=0$ at time $t=0$, in which
$m^{2}_{\phi}(T)$ changes sign and magnitude instantly, concluding
with the value $m_{\phi}^2(t)=-\mu^2$, $t>0$.

Beginning from this initial distribution, peaked around $\phi =
0$, we follow the evolution of the system, with Hamiltonian
determined from (\ref{S}). From the influence functional we define
the influence action $\delta A[\phi^+,\phi^-]$ by
$F[\phi^+,\phi^-] = \exp {i \delta A[\phi^+,\phi^-]}$. After
further defining $\Delta ={1\over{2}}(\phi^{+2} - \phi^{-2})$ and
$\Sigma ={1\over{2}}(\phi^{+2} + \phi^{-2})$, the real and
imaginary parts of the influence action are, in the one loop (two
vertices) and large $N$ approximations\footnote{The large-N
approximation singles out the two-vertex loop in the one-loop
perturbative approximation.},
\[
{\rm Re} \delta A = \frac{\bar g^2}{8} \int d^4 x\int d^4y ~
\Delta (x) K_{\rm q}(x-y) \Sigma (y),
\]
\[
{\rm Im} \delta A = - \frac{\bar g^2}{16} \int d^4x\int d^4y ~
\Delta (x) N_{\rm q} (x,y) \Delta (y),
\]
where $K_{\rm q} (x-y) = {\rm Im} G_{++}^2(x,y) \theta (y^0-x^0)$
is the dissipation kernel and $ N_{\rm q} (x-y) = {\rm Re}
G_{++}^2(x,y)$ is the noise (diffusion) kernel. $G_{++}$ is the
relevant closed-time-path correlator of the $\chi$-field at
temperature $T_0$.

The first step in the evaluation of the master equation is the
calculation of the density matrix propagator $J_{\rm r}$ from
Eq.(\ref{evolred}). In order to estimate the functional
integration which defines the reduced propagator, we perform a
saddle point approximation
\[
J_{\rm r}[\phi^+_{\rm f},\phi^-_{\rm f},t\vert\phi^+_{\rm
i},\phi^-_{\rm i}, t_0] \approx \exp{ i A[\phi^+_{\rm
cl},\phi^-_{\rm cl}]},
\]
where $A[\phi^+,\phi^-]= S[\phi^+]- S[\phi^-]+ \delta
A[\phi^+,\phi^-]$, and $\phi^\pm_{\rm cl}$ is the solution of the
equation of motion ${\delta Re
A\over\delta\phi^+}\vert_{\phi^+=\phi^-}=0$ with boundary
conditions $\phi^\pm_{\rm cl}(t_0)=\phi^\pm_{\rm i}$ and
$\phi^\pm_{\rm cl}(t)=\phi^\pm_{\rm f}$. It is very difficult to
solve this equation analytically. For simplicity, we assume that
the system-field contains only one Fourier mode with $\vec k =
\vec k_0$. We are motivated in this by the
observation\cite{boya,boyak0} that the exponentially growing
long-wavelengths increasingly bunch about a wave-number $k_0 <
\mu$,  which diminishes with time initially as $k_0^2 = {\cal
O}(\mu^2/t)$.

The classical solution is of the form
$\phi_{\rm cl}(\vec x, s) =f(s,t)\cos(\vec k_0 . \vec x)$
where $f(0,t)= \phi_{\rm i}$ and $f(t,t) = \phi_{\rm f}$.
Qualitatively, $f(s,t)$ grows exponentially with $s$ for $t\leq
t_{\rm sp}$, and oscillates for $ t_{\rm sp}<s<t$ when $t>t_{\rm
sp}$. We shall therefore approximate it, for $t\leq t_{\rm sp}$ as
\[
f(s,t)= \phi_{\rm f} {\sinh(\omega_0 s) \over {\sinh(\omega_0 t)}}
+ \phi_{\rm i} {\sinh[\omega_0 (t - s)] \over{\sinh(\omega_0
t)}}\,\, ,
\]
where $\omega_0^2 = \mu^2 - k_0^2$. In order to obtain the master
equation we must compute the final time derivative of the
propagator $J_{\rm r}$, and after that eliminate the dependence on
the initial field configurations $\phi^\pm_{\rm i}$ coming from
the classical solutions $\phi^\pm_{\rm cl}$ (see
Ref.\cite{lombmazz}).

To determine the onset of classical behaviour it is sufficient to
calculate just the correction to the normal unitary evolution
coming from the noise kernel. For clarity we drop the suffix ${\rm
f}$ on the final state fields. If $\Delta = (\phi^{+2} -
\phi^{-2})/2$ for the {\it final} field configurations, then the
relevant part of the  master equation for $\rho_{\rm
r}(\phi^+,\phi^-, t)$ is
\begin{equation}
i {\dot \rho}_{\rm r} = \langle \phi^+\vert [H,{\hat\rho}_{\rm r}]
\vert \phi^-\rangle - i\frac{\bar g^2}{16} V \Delta^2 D(k_0, t)
\rho_{\rm r}+ ... \label{master}
\end{equation}
The volume factor $V$ that appears in the master equation is due
to the fact we are considering a density matrix which is a
functional of two different field configurations, $\phi^\pm(\vec
x) = \phi^\pm \cos \vec k_0 . \vec x$, which are spread over all
space. The time-dependent diffusion coefficient $D(k_0,t)$ due to
each of the many external environmental fields
 is then given by

\begin{equation}
D_{\chi}(k_0, t)=
 \int_0^t ds ~ u(s)   \left [ {\rm Re}
G_{++}^2(2k_0; t-s) + 2 {\rm Re}G_{++}^2(0; t-s)\right].
\label{diff}
\end{equation}
In (\ref{diff}), $u(s) = \cosh^2\omega_0 s$ when $t \leq t_{\rm
sp}$, and is an oscillatory function of time when $t>t_{\rm sp}$.

Although $G_{++}$ is oscillatory at all times, for times $\mu t
>1$ (until the spinodal one) the exponential growth of $u(t)$
enforces a similar behaviour on $ D_{\chi}(k_0, t)$,
\begin{equation}
D_{\chi}(k_0, t) \sim {(k_{\rm B} T_0 )^2\over{\mu^3}} \omega_0~
\exp [2\omega_0 t], \label{unstD2}
\end{equation}
associated with the instability of the $k_0$ mode. For
long-wavelength modes $D_{\chi}(t) \sim (k_{\rm B} T_0/\mu)^2~
\exp [2\mu t]$. For $t>t_{\rm sp}$ the diffusion coefficient stops
growing, and oscillates around $D_{\chi}(k_0,t=t_{\rm sp})$.

In our present model the environment fields $\chi_{\rm a}$ are not
the only decohering agents. The environment is also constituted by
the short-wavelength modes of the self-interacting field $\phi$.
Therefore, we split the field as $\phi = \phi_< + \phi_>$, where
the system-field $\phi_<$ contains the unstable modes with
wavelengths longer than the critical value $\mu^{-1}$, while the
bath or environment-field contains the stable modes with
wavelengths shorter than $\mu^{-1}$ (in practice, whether the
separation is made at $k = \mu$ exactly or at $k\approx\mu$ is
immaterial\cite{Karra} by time $t_D$, when the power of the
$\phi$-field fluctuations is peaked at $k_0\ll\mu$). This gives an
additional contribution to the diffusion coefficient. Without the
additional powers of $N^{-1}$, a high temperature resummation of
loop diagrams \cite{parwani} is essential to get a reliable
$G_{++}$. This is beyond the scope of this letter. It will be
enough for our purposes to compute an upper bound on the
decoherence time $t_D$ only considering the external fields
$\chi_{\rm a}$. 

We estimate $t_{\rm D}$ by considering the approximate solution
(\ref{master}),\footnote{We are following the decoherence time
definition of \cite{zurekTA}.}
\[
 \rho_{\rm r}[\phi^+_<, \phi^-_<; t] \approx
\rho^{\rm u}_{\rm r}[\phi^+_<, \phi^-_<; t] ~ \exp
\bigg[-V\Gamma\int_0^t ds ~D(k_0, s) \bigg],
\]
where $\rho^{\rm u}_{\rm r}$ is the solution of the unitary part
of the master equation (i.e. without environment) and $D(k_0, s)$
denotes the total diffusion. It is obvious from this (and also
from (\ref{master})), that the diagonal density matrix just
evolves like the unitary matrix (the environment has almost no
effect on the diagonal part of $\rho_{\rm r}$). In terms of the
dimensionless fields $\bar\phi = (\phi_<^+ + \phi_<^-)/2\mu,$ and
$ \delta = (\phi_<^+ - \phi_<^-)/2\mu$, we have $\Gamma =
(1/16)\bar g^2 \mu^4\bar\phi^2\delta^2$.

 The system behaves classically
when the non-diagonal elements of the reduced density matrix are
much smaller than the diagonal ones. We therefore look at the
ratio
\begin{equation}
\left\vert \frac {\rho_{\rm r}[\bar\phi+\delta,\bar\phi-\delta;t]}
{\rho_{\rm r}[\bar\phi,\bar\phi;t]} \right\vert \approx \left\vert
\frac {\rho_{\rm r}^{\rm u}[\bar\phi+\delta, \bar\phi-\delta;t]}
{\rho_{\rm r}^{\rm u}[\bar\phi,\bar\phi;t]} \right\vert ~ \exp
\bigg[-V\Gamma\int_0^t ds ~D(k_0, s) \bigg ]\, .
\label{ratio}\end{equation}

It is not possible to obtain an analytic expression for the ratio
of unitary density matrices that appears in Eq.(\ref{ratio}). The
simplest approximation is to neglect the couplings of the system
field \cite{guthpi}. In this case the unitary density matrix
remains Gaussian at all times as

\begin{equation}\left\vert \frac {\rho_{\rm r}^{\rm u}[\bar\phi+\delta,
\bar\phi-\delta;t]} {\rho_{\rm r}^{\rm u}[\bar\phi,\bar\phi;t]}
\right\vert = \exp [-{T_{\rm c}\over{\mu}}\delta^2 p^{-1}(t)] ,
\label{uratio}
\end{equation}
where $p^{-1}(t)$, essentially $\mu^2
\langle\phi^2\rangle_t^{-1}$, decreases exponentially with time to
a value ${\cal O}(\lambda)$.   A full numerical calculation
\cite{diana} shows that $\rho_{\rm r}^{\rm u}$ becomes a
non-Gaussian function (the associated Wigner function becomes
non-positive). In any case, in the unitary part of the reduced
density matrix the non-diagonal terms are not suppressed.\footnote
{This should not be confused with the observation that the unitary
Gaussian density matrix does show classical correlation, whereby
the Wigner functional becomes localised in phase space about its
classical solutions. However, this has nothing to do with
eliminating quantum interference between different field
histories.}

The decoherence time $t_{\rm D}$ sets the scale after which we
have a classical system-field configuration. According to our
previous discussion, it can be defined as the solution to
\[
1\approx V\Gamma \int_{0}^{t_{D}} ds ~D(k_0, s)\geq V\Gamma
\int_{0}^{t_{D}} ds ~D_{\chi}(k_0, s).
\]
It corresponds to the time after which we are able to distinguish
between two different field amplitudes (inside a given volume
$V$).

Suppose we reduce the couplings ${\bar g}\sim\lambda$ of the
system $\phi$-field to its environment. Since, as a one-loop
construct, $\Gamma\propto{\bar g}^2\sim\lambda^2$, we might expect
that, as ${\bar g}$, $\lambda$ decrease, then $t_{\rm D}$
increases and the system takes longer to become classical.
Although this is the usual result for Brownian motion,
say\cite{zurek84}, it is not simply the case for quantum field
theory phase transitions. The reason is twofold. Firstly, there is
the effect that $\Gamma\propto T_0^2$, and $T_0^2
\propto\lambda^{-1}$ is non-perturbatively large for a phase
transition.  Secondly, because of the non-linear coupling to the
environment, obligatory for quantum field theory,
$\Gamma\propto{\bar\phi}^2$. The completion of the transition
finds ${\bar\phi}^2\simeq\eta^2\propto\lambda^{-1}$ also
non-perturbatively large. This suggests that $\Gamma$, and hence
$t_{\rm D}$, can be approximately independent of $\lambda$. In
fact, the situation is a little more complicated, but the end
result is that $t_{\rm D}$ does not increase (relative to $t_{\rm
sp}$) as the couplings become uniformly weaker.

In quantifying the decoherence time $V$ is understood as the
minimal volume inside which there is no possibility for coherent
superpositions of macroscopically distinguishable states for the
field (i.e. there is no `Schr\"odinger cat' states inside $V$).
Thus, our choice is that this volume factor is ${\cal
O}(\mu^{-3})$  since $\mu^{-1}$ (the Compton wavelength) sets the
smallest scale at which we need to look. In particular, $\mu^{-1}$
characterises the thickness of domain boundaries (walls) as the
field settles into its ground-state values. Inside this volume, we
do not discriminate between field amplitudes which differ by $
{\cal O}(\mu) $, and therefore we take $\delta \sim {\cal O}(1)$.
For $\bar\phi$ we set $\bar\phi^2\sim {\cal O}(\alpha /\lambda)$,
 where $\lambda\leq\alpha\leq 1$ is to be determined self-consistently
from the condition that, at time $t_{\rm D}$, $\langle
\phi^2\rangle_t\sim\alpha\eta^2$.

Note that the diagonalisation of $\rho_t$ occurs quickly, but not
so quickly that $\mu t \ll 1$.  Consequently, in order to evaluate
the decoherence time in our model, we have to use
Eq.(\ref{unstD2}). We obtain, for the upper bound on $t_D$,
\begin{equation}
\exp [2\mu t_{\rm D}] \approx {\lambda \sqrt{N}\bar g\over{\bar
g^2  \alpha}} = {\cal O}( \frac{\sqrt{N}}{\alpha}),
\end{equation}
whereby $\mu t_{\rm D}\simeq \ln (\eta /T_{\rm c}\sqrt{\alpha}))$.
The value of $\alpha$ is determined as $\alpha \simeq
\sqrt{\mu/T_{\rm c}}$.  For comparison, we find $t_{\rm sp}$, for
which $\langle \phi^2\rangle_t \sim
            \eta^2$, given by
            \begin{equation}
            \exp [2\mu t_{\rm sp}] \approx  {\cal O}( \frac{\eta^2}{\mu
            T_{\rm c}}).
            \end{equation}
The exponential factor, as always, arises from the growth of the
unstable long wavelength modes. The factor $T_{\rm c}^{-1}$ comes
from the $\coth(\beta\omega /2)$ factor that encodes the initial
Boltzmann distribution at temperature $T_0\gtrsim T_{\rm c}$. As a
result,
            \begin{equation}
            \mu t_{\rm sp} \sim
            \ln (\frac{\eta}{\sqrt{\mu T_{\rm c}}}),
             \label{tsp}
            \end{equation}
            whereby $1 < \mu t_D \leq \mu t_{\rm sp}$, with
            \begin{equation}
            \mu t_{sp} -\mu t_D\simeq \frac{1}{4}\ln (\frac{T_c}{\mu})>1,
            \label{dt}
            \end{equation}
            for weak enough coupling, or high enough initial
            temperatures.
Factors of ${\cal O}(1)$ have been omitted in the argument of the
logarithm in (\ref{dt}) and previous equations, requiring that
$T_c\gg\mu$.

 This is our main
result, that for the physically relevant modes (with small $k_0$)
classical behaviour has been established before the spinodal time,
when the ground states have became populated. We can say more in
that, for an instantaneous quench, nonlinear behaviour only
becomes important in an interval $\Delta t$, $\mu\Delta t = {\cal
O}(1)$, before the spinodal time\cite{Karra},
 and therefore
$\rho_{\rm r}$ becomes diagonal before non-linear terms could be
relevant. In this sense, classical behaviour has been achieved
before quantum effects could destroy the positivity of the Wigner
function $W_{\rm r}$.

\acknowledgments F.C.L. and F.D.M. were supported by Universidad
de Buenos Aires, CONICET, Fundaci\'on Antorchas and ANPCyT. R.J.R.
was supported, in part, by the COSLAB programme of the European
Science Foundation. We all thank the organisers of the Erice
Cosmology meeting of December, 2000, where some of this work was
performed.


\begin{references}
\bibitem{Kolb} E. Kolb and M. Turner, {\it The early Universe},
Addison Wesley (1990)

\bibitem{cormier} D. Cormier and R. Holman, Phys. Rev. {\bf D62}, 023520
(2000), and references therein

\bibitem{Kibble} T.W.B. Kibble, Phys. Rep. {\bf 67}, 183 (1980)

\bibitem{salman} S. Habib, Y. Kluger, E. Mottola, and J.P. Paz,
Phys. Rev. Lett. {\bf 76}, 4660 (1996)

\bibitem{mottola} F. Cooper, S. Habib, Y. Kluger, and E.
Mottola, Phys. Rev. {\bf D55}, 6471 (1997)

\bibitem{lombmazz} F.C. Lombardo and F.D. Mazzitelli, Phys. Rev. {\bf D53},
2001 (1996)

\bibitem{zurek84} W.H. Zurek, ``Reduction of the Wavepacket: How Long
Does it Take?'', lecture at tha NATO ASI {\it Frontiers of the
Nonequilibrium Statistical Mechanics}; Santa Fe, New Mexico; eds.
G. T.Moore and M. O. Scully (Plenum, New York, 1986)

\bibitem{habib} J.P. Paz, S. Habib, and W.H. Zurek, Phys. Rev. {\bf
D47}, 488 (1993).

\bibitem{diana} F.C. Lombardo, F.D. Mazzitelli, and D. Monteoliva, Phys. Rev.
{\bf D62}, 045016 (2000)

\bibitem{boya}  D. Boyanovsky, H.J. de Vega, and R. Holman, Phys. Rev.
{\bf D49}, 2769 (1994); D. Boyanovsky, H.J. de Vega, R. Holman,
D.-S. Lee, and A. Singh, Phys. Rev. {\bf D51}, 4419 (1995); S.A.
Ramsey and B.L. Hu, Phys. Rev. {\bf D56}, 661 (1997)


\bibitem{feynver} R. Feynman and F. Vernon, Ann. Phys. (N. Y.) 24, 118
(1963)

\bibitem{boyak0} D. Boyanovsky, D. Cormier, H.J. de Vega, R. Holman, and
S.P. Kumar, Phys. Rev. {\bf D57}, 2166 (1998)

\bibitem{zurekTA} W.H. Zurek, ``Prefered Sets of States, Predictability,
Classicality, and Environment-Induced Decoherence''; in {\it The
Physical Origin of Time Asymmetry}, ed. by J.J. Halliwell, J.
Perez Mercader, and W.H. Zurek (Cambridge University Press,
Cambridge, UK, 1994)

\bibitem{guthpi} A. Guth and S.Y. Pi, Phys. Rev. {\bf D32}, 1899 (1991); S.P.
Kim and C.H. Lee, hep-ph/0011227

\bibitem{Karra}G. Karra and R.J. Rivers, Phys. Lett. {\bf B414}, 28 (1997).

\bibitem{parwani} R.R. Parwani, Phys. Rev. {\bf D45}, 4695 (1992)
\end{references}
\end{document}